\documentclass[aps,pra,showpacs]{revtex4}
\usepackage{amsbsy,latexsym}
\usepackage{amsfonts}
\usepackage{amssymb}
\usepackage[mathscr]{eucal}

\newcommand{\ket}[1]{\vert #1 \rangle}
\newcommand{\bra}[1]{\langle #1 \vert}
\newcommand{\Bb}{{\boldsymbol B}}
\newcommand{\bb}{{\boldsymbol b}}
\newcommand{\Ab}{{\boldsymbol A}}
\newcommand{\Sb}{{\boldsymbol S}}
\newcommand{\Nb}{{\boldsymbol N}}
\newcommand{\D}[3]{\mathfrak D^{(#1)}_{#2 #3}}
\newcommand{\Dc}[3]{\mathfrak D^{(#1) *}_{#2 #3}}
\newcommand{\Da}[4]{\mathfrak D^{(#1,#2)}_{#3 #4}}
\newcommand{\Dac}[4]{\mathfrak D^{(#1,#2)*}_{#3 #4}}
\newcommand{\U}[1]{U^{(#1)}}
\begin{document}
\title{Optimal encoding and decoding of a spin direction}
\author{E.~Bagan, M.~Baig, A.~Brey and R.~Mu\~{n}oz-Tapia}
\affiliation{Grup de F\'{\i}sica Te\`{o}rica \& IFAE, Facultat de Ci\`{e}ncies,
Edifici Cn, Universitat Aut\`{o}noma de Barcelona, 08193 Bellaterra
(Barcelona) Spain}
\author{R.~Tarrach}
\affiliation{Departament d'Estructura i Constituents de la
Mat\`{e}ria, Universitat de Barcelona, Diagonal 647, 08028 Barcelona,
Spain}
\date{December 1, 2000}

\begin{abstract}
For a system of $N$ spins 1/2 there are quantum states that
can encode a direction in
an intrinsic way. Information on this direction can later be
decoded by means of a
quantum measurement.
We present here the optimal encoding and decoding procedure using the
fidelity as a figure of merit. We compute the maximal fidelity and
prove
that it is directly related to the largest
zeroes of the Legendre and Jacobi polynomials. We show that this
maximal
fidelity approaches unity quadratically in $1/N$.
We also discuss this result in terms of the dimension of the encoding
Hilbert space.
\end{abstract}


\maketitle

\section{Introduction}\label{sect-introduction}
Entanglement and superposition are the most characteristic features of
quantum states. They play a central role in the storage and
transmission of information in the quantum world and are responsible
of the many remarkable, and often intriguing, quantum effects that are
constantly being discovered.
These effects, in turn, are providing new
insights in the difficult task of understanding
quantum information.

Some time ago Peres an Wooters~\cite{pw} posed an interesting
question. Imagine a quantum system composed of several
subsystems, which are not necessarily entangled.  How can we
learn more about this system? by performing measurements
on the individual subsystems or on the
system as a whole? They
showed evidence that the latter, the so-called collective
measurements, are more informative. Obviously entanglement is the property
responsible for this. In this case, however, it is not explicit, since the
system can be chosen to be in a product state, but hidden in the
collective measurement.

Later Massar and Popescu~\cite{mp} addressed a more concrete
problem. Imagine Alice has a system of $N$ parallel spins. She
can use this system to tell Bob the direction along which some
given unit vector $\vec n$ is pointing. She just have to rotate,
or prepare in some other way, the state of her system so that it
becomes an eigenstate of $\vec n\cdot \vec S$, the projection of
the total spin in the $\vec n$ direction. The state is then sent
to Bob whose task is to determine the direction encoded in the
state. He will need to perform a collective measurement and from
each one of its outcomes, labeled with an index~$r$, he will have
a guess for Alice's direction given by a unit vector $\vec n_{r}$.
To quantify the quality of Bob's measurement Massar and Popescu
used the average fidelity which is defined by $F= \sum_{r} \int
dn\,  (1+\vec n\cdot\vec n_{r})/2 \;P_{r}(\vec n) $, where
Alice's directions $\vec n$ are assumed to come from an isotropic
source. Here $P_{r}(\vec n)$ is the probability of getting the
outcome $r$ if Alice's direction is $\vec n$, and $dn$ is the
rotationally invariant measure on the unit 2-sphere. The authors
proved that the maximal average fidelity Bob can achieve is
$F=(N+1)/(N+2)$ which is readily seen to approach unity linearly:
$F\sim 1-1/N$. Explicit realizations of the optimal measurements
with a finite number of outcomes were obtained in~\cite{derka}
for arbitrary $N$ and minimal versions of these measurements for
$N$ up to seven are in~\cite{lpt}.

A new
surprise was recently presented in~\cite{gp}. In this paper the
authors consider $N=2$ and show that states with two
anti-parallel spins, $\ket{\uparrow\downarrow}$, $\ket{\downarrow\uparrow}$
provide a better encoding of Alice's directions
than the two parallel spin states used in~\cite{mp,derka,lpt}.
The average fidelity is now $(3+\sqrt{3})/6$ which is larger than $3/4$ for
two parallel spins, i.e., Bob can have a better determination of
Alice's direction
if she uses anti-parallel spins.
This is a startling result, since classically one would expect that a
direction is encoded equally as well in a state pointing one way
as in one pointing the opposite way. The main reason why this is not
so in the quantum world, as it will become
clear from
our  work, is the different dimensionality of the Hilbert spaces to which
two parallel or two anti-parallel spin states belong .

At this point, the obvious reaction is to ask ourselves
what are the best states Alice can use to encode directions. Since the
very natural state with only parallel spins
is not optimal for $N=2$, we expect that neither will it be
for arbitrary $N$. Hence, one has to search the optimal encoding
state among all the eigenstates of $\vec n\cdot \vec S$ that Alice can
produce.
In this paper we present this very general analysis.
We compute the maximal average fidelity (hereafter we will usually drop
`average' when there is no ambiguity) for arbitrary $N$ and show that it
approaches unity quadratically in $1/N$, as compared to the
linear behavior found in~\cite{mp} for
parallel spins.
As a byproduct, we also compute the~maximal fidelity for encoding
states of two arbitrary spins $s$ such as two nuclei.
A short description of the main results of this
analysis was presented in \cite{us}.
These results have been recently corroborated by numerical analysis~\cite{ps}.

The paper is organized as follows. In section~\ref{sect-2 spins} we
introduce our notation and conventions and present
a detailed calculation of the maximal fidelity for $N=2$.
We show that the fidelity obtained by Gisin and
Popescu in~\cite{gp} is optimal (a result also
obtained in~\cite{massar} using different methods). In
section~\ref{sect-2 spins s} we analyze the more general case of
two spin $s$ states.
The analysis for any number $N$ of spins is in
section~\ref{sect-general case} and our
results and discussion are in section~\ref{sect-results and discussion}.
We conclude
with an appendix containing technical details.

\section{Two spins}\label{sect-2 spins}

We start by assuming that
Alice has two spins in a general eigenstate of $\vec n\cdot S$
(We skip the analysis of the simplest
situation in which Alice has only a spin.
The reader can find it
in~\cite{mp,us}, and our general
formulae of section~\ref{sect-general case}
can be also particularized to this case).
We can think of it as a fixed eigenstate
of $S_{z}=\vec z\cdot \vec S$ ($\vec z$ is the unit vector
pointing along the $z$ direction) that Alice
has rotated into the direction
$\vec n=(\cos\phi\sin\theta,\sin\phi\sin\theta,\cos\theta)$.
It is convenient to work in the irreducible representations of
$SU(2)$. In the present case, {\boldmath$1/2\otimes 1/2= 1\oplus 0$},
the general form of
this fixed eigenstate is
\begin{equation}\label{A2}
|A\rangle=A_{+}|1,1\rangle
+A_0|1,0\rangle+A_-|1,-1\rangle+A_s|0,0\rangle,
\end{equation}
where, as usual, the normalized states of the basis, $\ket{j,m}$, are  labeled
by the total
spin $S$ and the third component $S_z$:
$S{}^2|j,m\rangle=j(j+1)|j,m \rangle$ and $ S_z |j,m\rangle
=m|j,m\rangle$. In the following we stick to the general form
(\ref{A2}) to treat all the cases jointly, but one should keep in
mind that only combinations with definite $S_z$  will be
relevant for our analysis. The rotated state $U(\vec n)|A\rangle$, where
$U(\vec n)$ is the
element of the $SU(2)$ group associated to the rotation
$\vec{z}\to \vec{n}=R\;\vec{z}$, is precisely Alice's general eigenstate
of $\vec n\cdot\vec S$. Obviously $U(\vec n)$ is reducible since it has the
form $U(\vec n)=\U1 (\vec n) \oplus \U0 (\vec n)$, where $\U j $
denotes the $SU(2)$ irreducible
representation of spin $j$.

Next, Alice  sends the rotated state to Bob, who tries to determine
$\vec n$ from his measurements.
The most general one he can perform  is a positive
operator valued measurement (POVM).
We specify this POVM by giving
a set of
positive Hermitian operators $\{O_{r}\}$, that are a resolution
of the identity
\begin{equation}
 \mathbb I= \sum_{r} O_{r}
 \label{povm}.
\end{equation}
For each outcome, $r$, Bob makes a guess, $\vec n_{r}$, for the
direction. As we brought up in the introduction,
the quality of the guess is quantified in terms of the fidelity which
we
can view as a `score'. To Bob's guess $\vec n_{r}$, we give the `score'
$f=(1+\vec{n}\cdot\vec{n}_r)/2$. We see that he fidelity $f$ is unity if Bob's
guess coincides with Alice's direction and it is zero when they are
opposite. Thus, if $\vec n$ is isotropically distributed the
average fidelity can be written as
\begin{equation}
F=\sum_r \int dn{1+\vec{n}\cdot\vec{n}_r\over 2}
\mathrm{tr}[\rho(\vec{n}) O_r], \label{fidelity}
\end{equation}
where $\rho(\vec{n})=U(\vec{n})|A\rangle\langle A|U^{\dag}(\vec{n})$ and
$dn$ was defined in the introduction.
The
evaluation of $F$ can be greatly simplified by exploiting the
rotational invariance of the integral (\ref{fidelity}). If we
define $R_r$ through the relation
\begin{equation}\label{nr}
\vec{n}_r=R_r \vec{z}
\end{equation}
and make the change of variables
\begin{equation}\label{n'}
    R_r^{-1}\vec{n}\to\vec{n}.
\end{equation}
We have
\begin{equation}
F=\sum_r \int d{n} \frac{1+\vec{n}\cdot
\vec{z}}{2}\mathrm{tr}[\rho(\vec{n}) \Omega_r],
\label{fidelity-1}
\end{equation}
where
\begin{equation}\label{omega-r}
\Omega_r=U^{\dag}(\vec n_r) O_r U(\vec n_r).
\end{equation}

Notice that
in general $\sum_r\Omega_r \neq \mathbb I$.
We can regard $\Omega_{r}$ as fixed or reference projectors
associated to the single direction $\vec z$. In this sense, they are the
counterpart of Alice's fixed state $\ket{A}$.
Inserting four times
the closure relation $\sum_{k}\ket{k}\bra{k}=\mathbb I$, where
$k=+,0,-,s$, and
$\{\ket{k}\}$ is the  basis of the representations
{\boldmath$1\oplus 0$},
\begin{eqnarray}\label{basis}
 \ket{\pm}&=&\ket{1,\pm 1} \nonumber \\
\ket{0}&=&\ket{1,0} \nonumber \\
\ket{s}&=&\ket{0,0},
\end{eqnarray}
we  obtain
\begin{equation}
F=\sum_{kijl}A_i^*A_l\, \omega_{kj} \int {dn
\frac{1+\cos\theta}{2}\mathfrak D_{ki}^* (\vec n)\mathfrak D_{jl}}(\vec n).
 \label{fidelity-2}
\end{equation}
Here the indices $k$, $i$, $j$, $l$, run also over $+,0,-,s$;
$\mathfrak D_{kj}(\vec
n)=[\mathfrak D^{(1)}\oplus
\mathfrak D^{(0)}]_{kj}(\vec n)=\bra{k}U(\vec n)\ket{j}$ are
the $SU(2)$ rotation matrices  in the
{\boldmath$1\oplus 0$} representations, and
\begin{equation}\label{omegas}
\omega_{kj}=\sum_r \langle k|\Omega_r|j\rangle.
\end{equation}
Now, one can easily evaluate the integrals and obtain the
fidelity
\begin{equation}
F=\mathsf{A}^{\dag} \mathsf{W}\mathsf{A}, \label{awa}
\end{equation}
where $\mathsf{A}=(A_+, A_0, A_-, A_s)^{t}$ and
$\mathsf{A}^{\dag}$ is its transposed complex conjugate. The
matrix $\mathsf{W}$ is
\begin{equation}\label{matrix-omega}
\mathsf{W}= \pmatrix{\displaystyle
\frac{3\omega_{++}+2\omega_{00}+\omega_{--}}{12} & * & * & * \cr
   * &\displaystyle \frac{\omega_{++}+\omega_{00}+\omega_{--}}{6} & * &
   \displaystyle\frac{
\omega_{0s}}{6}
   \cr
   * & * &\displaystyle \frac{\omega_{++}+2\omega_{00}+3\omega_{--}}{12} & *
   \cr
   * & \displaystyle\frac{ \omega_{s0}}{6} & * &\displaystyle
   \frac{\omega_{ss}}{2}} ,
\end{equation}
where the entries marked with $*$ are not relevant for our analysis
since we only consider eigenstates of $S_{z}$ for the fixed
states $\ket{A}$. These, and the corresponding rotated states $U(\vec
n)\ket{A}$,
are the only ones that point along a
definite direction in an absolute sense, i.e., even if Alice and
Bob do not share a common reference frame.
>From its definition~(\ref{omegas}), it follows that $\omega_{jj}$ are
real nonnegative numbers but
$\omega_{ij}$ are in general complex numbers for $i\neq j$.
There are other constrains on $\omega_{ij}$ stemming from
the condition $\sum_r O_r= \mathbb I$:
\begin{equation}\label{condition-1}
  \omega_{ss}=1, \ \ \ \ \ \ \ \sum_{l=+,0,-}\omega_{ll}=3.
\end{equation}
Because of the Schwarz inequality, we also have
\begin{equation}\label{condition-2}
|\omega_{0s}|^2\leq \omega_{00}\omega_{ss}=\omega_{00}.
\end{equation}
Let us discuss the implications of these equations
for
different values of $m$.
\bigskip

\noindent{\bf Case \boldmath{$m=\pm 1$}}
\medskip

The fixed state $\ket{A}$ for $m=1$ is simply $\ket{A}=\ket{1,1}$, i.e.,
$A_{+}=1$ and $A_{0}=A_{-}=A_{s}=0 $. In this case the fidelity
is given by the element $\mathsf W_{++}$ of (\ref{matrix-omega}),
\begin{equation}
F=\mathsf W_{++}=\frac{3\omega_{++}+2\omega_{00}+\omega_{--}}{12}=\frac{3}{4}-
\frac{\omega_{00}+2\omega_{--}}{12}\leq \frac{3}{4},
\end{equation}
where the second condition in~(\ref{condition-1}) has been used.
The maximal value, which we denote by $F_+$, is then
\begin{equation}
F_+=\frac{3}{4}.
\end{equation}
This value occurs for
\begin{equation}
\label{optimal condition}
\omega_{--}=\omega_{00}=0\qquad\Rightarrow\qquad
\omega_{++}=3.
\end{equation}
The
case $m=-1$, for which $\ket{A}=\ket{1,-1}$, is completely analogous
with the index  substitution $+\leftrightarrow
-$. The maximal value of the fidelity is also $F_{-}=3/4$.
\bigskip

\noindent{\bf Case \boldmath{$m=0$}}
\medskip

For $m=0$ one has
$\ket{A}=A_0\ket{1,0}+A_s\ket{0,0}$, with $|A_0|^2+|A_s|^2=1$.
The maximal fidelity is the largest eigenvalue of the
$2\times2$ submatrix of
(\ref{matrix-omega}) corresponding to the $m=0$ subspace:
\begin{equation}\label{eigenvalue}
F= \frac{3+|\omega_{0s}|}{6}\leq\frac{3+\sqrt{\omega_{00}}}{6}.
\end{equation}
It reaches its maximal value, $F_0$, for
\begin{equation}\label{omega-condition}
\omega_{00}=3\quad\Rightarrow\quad \omega_{++}=\omega_{--}=0.
\end{equation}
Substituting back into~(\ref{eigenvalue}) we obtain  \cite{gp}
\begin{equation}\label{fidelity-2x2}
 F_0=\frac{3+\sqrt{3}}{6}.
\end{equation}
The corresponding eigenvector is
\begin{equation}\label{eigenvector-2}
  |A\rangle=\frac{1}{\sqrt{2}}|1,0\rangle+\frac{e^{i
  \delta}}{\sqrt{2}}|0,0\rangle,
\end{equation}
where the phase is the unconstrained parameter
$\delta=\arg{\omega_{s0}}$. Notice that the family of states
(\ref{eigenvector-2}) contains entangled as well as unentangled states.
With the choice $e^{i \delta}=\pm 1$ one obtains the product states
$\ket{\uparrow\downarrow}$,
$\ket{\downarrow\uparrow}$; precisely those
considered by Gisin and Popescu~\cite{gp}, which led
them to the conclusion that anti-parallel spins are better than
parallel spins for encoding
a direction.

>From this analysis one can also obtain important information
about the optimal POVM. Taking into account that
one can always take the projectors $O_r$ to be one-dimensional~\cite{davies},
we can write Bob's reference projectors $\Omega_{r}$ as
\begin{equation}\label{psi-r}
 \Omega_r=c_r \ket{\Psi_r}\bra{\Psi_r},
\end{equation}
where $\ket{\Psi_r}$ are normalized states and $c_r$ are
positive numbers. The values of $\omega_{ij}$ (see
Eq.~\ref{omegas})  endow the information about the components of
$\ket{\Psi_r}$ in the spherical basis~(\ref{basis}). To be specific,
consider states with $m=0$. The maximal fidelity
condition~(\ref{omega-condition})
implies that the states $\ket{\Psi_r}$ must
have also $m=0$, hence $\ket{\Psi_r}=\alpha_r
\ket{1,0}+\beta_r\ket{0,0}$. This result is, to some extent, what one
expects: In order for a
POVM to be optimal, the measurement must project on states
as similar as possible to the signal state. Further, the Schwarz
inequality~(\ref{condition-2}) becomes equality if and only if
$\alpha_r/\beta_r=\lambda$ for all $r$. If this is the case, the
fidelity can reach the maximal value $F_{0}$.  Then, imposing the POVM
conditions~(\ref{condition-1}) it is straightforward to verify
that all $\ket{\Psi_r}$ must coincide with a single state, which we
denote by $\ket{B}$,
\begin{equation}\label{B-2}
  \ket{\Psi_r}=\ket{B}=\frac{\sqrt{3}}{2}\left| 1,0 \right\rangle
+\frac{
 {\rm e}_{{}}^{i\delta }}{2}\left| 0,0\right\rangle,
\end{equation}
The relative weights of the $\ket{1,0}$ and $\ket{0,0}$
components, $\sqrt{3}:1$, are easily understood as being the
square root of the dimension of the Hilbert spaces corresponding to $j=1$ and $j=0$. We
therefore see that optimal POVMs can be obtained by rotating  the
single reference state $|B\rangle$.
The weights $c_r$ are free parameters except for the constrain
\begin{equation}
 \sum_{r} c_{r}=4.
 \label{constrain}
\end{equation}

Because the Hilbert space has dimension four, a POVM (optimal or
not) must consist of at least four projectors. Let us show that
indeed an optimal POVM with this minimal number of projectors
exists. Since the number of projectors in the POVM equals the
dimension of the Hilbert space, we are actually dealing with a
von Neumann measurement, i.e.,
\begin{equation}
O_r O_s=O_r\delta_{rs}.
\end{equation}
Hence,  $ \langle\Psi_r|\Psi_r\rangle=1 \Rightarrow c_{r}=1$
for the four values of
$r$, which is, of course, consistent with~(\ref{constrain}).
Inverting (\ref{omega-r}) and
taking into account (\ref{psi-r}),
we see that the four unit vectors $\vec n_{r}$ have to be chosen so that
\begin{equation}\label{povm-2x2}
 \sum_{r=1}^{4}O_r=\sum_{r=1}^{4}U(\vec n_r)\ket{B}
 \bra{B}U^{\dagger}(\vec n_r)
=\mathbb I.
\end{equation}
By symmetry, they should
correspond to the vertices of a tetrahedron inscribed in a unit sphere, i.e.,
$\vec n_{r}=
(\cos\phi_{r}\sin\theta_{r},\sin\phi_{r}\sin\theta_{r},\cos\theta_{r})$
with
\begin{equation}
\begin{array}{rclcrclcrcl}
\cos\theta_1&=&1,& \qquad &\phi_1&=&0; & \qquad & &&\\
\cos\theta_r&=&-\frac{1}{3},&&\phi_r&=&(r-2)\frac{2\pi}{3},&&
r&=&2,3,4.
\end{array}
\label{tetrahedron}
\end{equation}
It is easy to verify that with this choice condition~(\ref{povm-2x2})
is fulfilled and the maximal fidelity~(\ref{fidelity-2x2}) is
attained.
One can check that the four projectors~(\ref{povm-2x2}) are equal to those
already considered by Gisin and Popescu in~\cite{gp}. Our aim here was just to
present a
motivated explanation
for their choice of POVM.
Finite optimal POVMs for $N>2$ are less straightforward to
obtain. However, the results of~\cite{derka,lpt}, which enables us to
construct finite POVMs for
code states with maximal $m$,
$\ket{N/2,N/2}=\ket{\uparrow\uparrow\ \stackrel{\mbox{\tiny $N$)}}{\dots}
\ \uparrow}$, can also be used
here for
other values of~$m$. We will comment on this issue in our last
section.

\bigskip

After dwelling on minimal POVMs, it is convenient to consider also the
other end of the spectrum: POVMs with infinitely many outcomes or
continuous POVMs~\cite{holevo}. They will be used in the general
analysis in the sections
below, where they will prove very efficient.
Recall that for any finite measurement on
isotropic distributions it is always possible to find a continuous
POVM that gives the same fidelity~\cite{derka}. Therefore,
restricting ourselves to this type of measurements do not imply any loss of
generality.
We illustrate this point for $N=2$ and $m=0$
to introduce the notation that will be used in the following sections.

We have seen that the matrix elements
$\omega_{ij}$ contain all the information
required for
computing
the fidelity, independently of any particular choice of POVM. Any
measurement
for which $\omega_{ij}$ satisfy the condition~(\ref{optimal
condition}) for $m=1$ or~(\ref{omega-condition}) for $m=0$ is surely
optimal. A continuous POVM is just a particularly simple and useful
realization. It amounts to taking the
index $r$ to be continuous, i.e.,
\begin{equation}\label{int-c}
\sum_r \to \int d n_B,
\end{equation}
where
the
subindex $B$ in the invariant measure refers to Bob (measuring device).
Substituting~(\ref{psi-r})
into~(\ref{omegas})
one obtains in the continuous
version
\begin{equation}
\omega_{kj}=\int d n_B\,  c(\vec n_B)\, \langle k|B\rangle\langle
B|j\rangle,
\end{equation}
where $\ket{B}$ is the normalized state~(\ref{B-2}) and~$c(\vec n_B)$ is a
continuous positive weight, which plays the role of~$c_{r}$ and
according to~(\ref{constrain})
must satisfy
\begin{equation}
\int d n_B\, c(\vec n_B) = 4.
\end{equation}
We now show that in fact $c(\vec n_B)$ is  a constant and, hence, equal
to~4.
Condition~(\ref{povm-2x2})
reads
\begin{equation}\label{povm-c}
 \int d n_B\,  c(\vec n_B)\, U(\vec n_{B})\ket{B}
 \bra{B}U^{\dagger}(\vec n_{B})
=\mathbb I,
\end{equation}
which is equivalent to
\begin{equation}\label{norm-2-c}
  \frac{2j+1}{4}\int d n_B\,  c(\vec n_B)\, \D j m 0 (\vec n_{B})\;
      \Dc {j'} {m'} 0 (\vec n_{B}) =
      \delta_{j j'}\delta_{m m'}, \qquad j,j'=0,1.
\end{equation}
Using the well known orthogonality relation of the matrix
representations
of $SU(2)$~\cite{wktung},
\begin{equation}
 \int dn\;
  \D {j} {m_1} {m_2} (\vec n)\;
 \Dc {j'} {m_1'} {m_2} (\vec n)
  =\frac{1}{2j+1}\delta_{j j'}\delta_{m_1\,m_1'},
 \label{su2-2d}
\end{equation}
one obtains
\begin{equation}
c(\vec n_B)\equiv c= 4,
\end{equation}
which is just the total dimension ($3+1$) of the Hilbert space to
which the state~(\ref{B-2}) belongs. Therefore, the projectors $O(\vec
n_{B})=c \, U(\vec n_{B})\ket{B}
 \bra{B}U^{\dagger}(\vec n_{B})$ in~(\ref{povm-c}) describe an optimal
continuous
POVM. They are obtained from the fixed
state~(\ref{B-2}) in a manner analogous to the construction of the minimal
POVM in~(\ref{povm-2x2}) and~(\ref{tetrahedron}), excepting the
constant factor~$c$ required by
the normalization of the matrix representations of~$SU(2)$.

\bigskip

To complete the analysis of $N=2$, we calculate the maximal fidelity for
a given (non-optimal) fixed state $\ket{A}$ with~$m=0$. Without any loss of
generality it can be written as
\begin{equation}\label{fixed-state}
\ket{A}=|A_0|\; \ket{1,0}+ |A_s|{\rm e}^{i\delta}\; \ket{0,0},\qquad
|A_{0}|^2+|A_{s}|^2=1;
\end{equation}
where we have used the same phase convention as in~(\ref{eigenvector-2}).
From~(\ref{matrix-omega}), and the constrains~(\ref{condition-1})
and~(\ref{condition-2}), it is straightforward to see
that the maximal value of the fidelity is
\begin{equation}
 F_A=\frac{1}{2}+\frac{|A_0||A_s|}{\sqrt{3}}.
 \label{fidelity-fixed}
\end{equation}
To attain this value, Bob must perform an optimal POVM, characterized
by~(\ref{B-2}). He may use, for instance, the minimal
one (Eqs.~\ref{povm-2x2}--\ref{tetrahedron}), or the continuous one,
$O(\vec n_{B})$.
This result shows that for any fixed state~(\ref{fixed-state})  with
$1/2<|A_0|<\sqrt{3}/2$ the fidelity
is higher than that of the parallel case (i.e., $m=\pm 1$) for
which $F=F_{\pm}=3/4$.

\section{Two spins $\Sb$}\label{sect-2 spins s}

Imagine now that instead of two spin-$1/2$ Alice can use
two equal arbitrary spins
$s_1=s_2=s$ to encode the directions. This
can be seen as a generalization of the simple case studied
in the preceding section.  However the most important feature of this
analysis, as it will be
shown in section~\ref{sect-general case},
is that it provides the solution of our original problem, namely,
that of obtaining the maximal fidelity when
Alice has $N$ spin-1/2 particles at her disposal.

According to the Clebsch-Gordan decomposition, a normalized
eigenvector of the total spin in the $z$-direction with eigenvalue
$m_{A}$ can be
written
as
\begin{equation}\label{A-general}
\ket{A}=\sum_{j=m_A}^{J} A_{j}\ket{j,m_A};\qquad \sum_{j=m_{A}}^{J}
|A_{j}|^2=1;
\end{equation}
where $J=2s$. The state $\ket{A}$ and its components $A_{j}$
should carry the label $m_A$ to denote the different
eigenvalues of $S_{z}$, however, we will drop it  to simplify
the notation. A general eigenstate of $\vec n\cdot\vec S$ has the form
$U(\vec n)\ket{A}$, where $U(\vec n)$ is now
\begin{equation}
U(\vec n)=\bigoplus_{j=m_{A}}^{J}
U^{(j)}(\vec n)
\end{equation}
The POVM projectors can be constructed from a fixed state $\ket{B}$
of the form
\begin{equation}
 \ket{B}=
 \sum_{j=m_B}^{J} B_j\ket{j,m_B},
 \label{B-general}
\end{equation}
namely,
$
O(\vec n_{B})= c\, U(\vec n_{B})\ket{B}\bra{B}U^\dagger(\vec n_{B})
$.
Note that $\ket{B}$ is an eigenvector of $S_{z}$ with eigenvalue
$m_{B}$, though
we also drop the label~$m_B$ here. The absolute value of the
coefficients $B_j$ and the positive
weight~$c$ are
determined by the completeness relation $\int d n_B\; O(\vec n_B) = \mathbb
I$, which using~(\ref{su2-2d})
leads to the normalization condition
\begin{equation}\label{B-norm}
  |B_j|=\sqrt{\frac{2j+1}{c}},
\end{equation}
and a value for $c$ given by
\begin{equation}
c=(J+1)^2-m_B^2.
\end{equation}
Notice that the factor~$2j+1$ in~(\ref{B-norm}) is just the
dimension of the Hilbert space of the irreducible
representation~{\boldmath $j$}
of $SU(2)$,  and~$c$ is the dimension of the total Hilbert space.
Thus, (\ref{B-general}) is the straight generalization of the
states~(\ref{B-2}).
The fidelity can be written as
\begin{equation}
F=c\sum_{j,j'=m}^J A_j A^*_{j'} B_j^* B_{j'}\int dn
\frac{1+\cos\theta}{2} \D {j} {m_B} {m_A} (\vec n)\,\Dc {j'} {m_B} {m_A}
(\vec n),
\label{fidelity-general-1}
\end{equation}
where
\begin{equation}\label{m}
m=\max(m_A,m_B).
\end{equation}
The integral in~(\ref{fidelity-general-1}) can be easily computed by
noticing that $\cos\theta=\D 1 0 0 (\vec n)$. Using again the orthogonality
relations~(\ref{su2-2d}) we have
\begin{equation}\label{su2-3d}
\int dn \;\cos\theta\;
  \D {j} {m_1} {m_2} (\vec n)\;
 \Dc {j'} {m_1'} {m_2} (\vec n)
  =\frac{1}{2j'+1}
  \langle 10;j m_1|j'm_1'\rangle\langle 10;j
  m_2|j'm_2\rangle,
\end{equation}
were $\langle j_1 m_1;j_2 m_2|j_3 m_3\rangle$ are the
Clebsch-Gordan coefficients of {\boldmath
$j_1\otimes j_2\to
j_3$}.
The fidelity can be cast as
\begin{equation}\label{fidelity-general-2}
F=\frac{1}{2}+\frac{1}{2}\sum_{j=m}^{J}  \mu_j |A_j|^2 +
\frac{1}{2}\sum_{j=m+1}^{J} \left( A_{j-1}^* A_{j}
\nu_j^* + A_{j-1} A_{j}^* \nu_j \right )-{1\over2}\sum_{j=m_A}^{m-1} |A_j|^2,
\end{equation}
where the last term is zero for $m_A<m_B$ and the
coefficients $\mu_j$ and $\nu_j$ are
\begin{eqnarray}
\mu_j&=&\frac{m_A m_B}{j(j+1)}\label{muj}\\
\nu_j&=&\frac{e^{i\delta_j}}{j}\left(\frac{(j^2-m_A^2)
(j^2-m_B^2)}{4 j^2-1}\right)^{1/2} \label{nuj}.
\end{eqnarray}
The phases~$\delta_j$ in~(\ref{nuj}) are arbitrary. They are just the
generalization of the single free phase of~(\ref{B-2}). Here we
have $\delta_j=\arg(B_j^* B_{j-1})$. The maximal fidelity is
achieved by choosing $\delta_{j}$ equal to the phases
of the
signal state~$\ket{A}$:
\begin{equation}
\delta_j=\arg(B_j^* B_{j-1})=\arg(A_j^* A_{j-1}). \label{phases}
\end{equation}
We see now that all terms in~(\ref{fidelity-general-2}) are explicitly positive
with the exception of the last one, which necessarily vanishes for
optimal states~$\ket{A}$, i.e., $A_j=0$ for $j<m$. Gathering all these
results, we obtain for the fidelity:
\begin{equation}\label{fidelity-general-3}
F=\frac{1}{2}+\frac{1}{2} \mathsf{A}^t \mathsf{M}  \mathsf{A}.
\end{equation}
Here
$\mathsf{A}^t=(|A_{J}|,|A_{J-1}|,|A_{J-2}|,\dots)$
is the transpose of $\mathsf A$, and
$\mathsf{M}$ is a real matrix of tridiagonal form
\begin{equation}\label{matrix}
\mathsf{M}=\pmatrix{d_{l}&c_{l-1}&
         &        &       \cr
                    c_{l-1}&\ddots&\ddots
    &\phantom{\ddots} \raisebox{2.0ex}[1.5ex][0ex]{\LARGE
0}\hspace{-.5cm}       &       \cr
                         &\ddots&d_{3}    &c_{2}  &       \cr
                         & \phantom{\ddots}    &c_{2}&d_{2}&c_{1}
\cr \hspace{.5cm} \raisebox{2.0ex}[1.5ex][0ex]{\LARGE
0}\hspace{-.5cm} &&\phantom{\ddots}&c_{1}&d_{1}}
\end{equation}
with
\begin{equation}
l=J+1-m , \label{n-N}
\end{equation}
and
\begin{eqnarray}
d_k&=&\mu_{k+m-1} \nonumber \\
c_k&=&|\nu_{k+m}|\label{dk-ck}.
\end{eqnarray}
The largest eigenvalue, $x_{l}$, of $\mathsf M$ determines the
maximal fidelity through the relation
\begin{equation}\label{fidelity-general-4}
  F=\frac{1+x_l}{2} .
\end{equation}
To find $x_{l}$, we set up a recursion relation for  the
characteristic polynomial of $\mathsf M$:
\begin{equation}\label{ch-p}
Q_{l+1}(x)=(d_{l+1}-x) Q_{l}(x) -c_{l}^2 Q_{l-1}(x),
\end{equation}
with the starting values $Q_{-1}(x)=0$ and $Q_0(x)=1$.
Eq.~\ref{ch-p} resembles the recursion relation of orthogonal
polynomials, but at first sight the solution does not seems
straightforward at all. We thus work out in detail the simplest
case for which $m_A=m_B=0$. For this particular instance~(\ref{ch-p}) reads
\begin{equation}
Q_{l+1}(x)=-x Q_{l}(x)-\frac{l^2}{4
l^2-1}Q_{l-1}(x),\label{legendre-1}
\end{equation}
where we have used the definitions (\ref{muj}), (\ref{nuj}) and (\ref{dk-ck}).
We can rewrite~(\ref{legendre-1}) as
\begin{equation}
(l+1)\left[ -\frac{(2l+1)(2l-1)}{(l+1)l} Q_{l+1}(x)\right]=
(2l+1)x \left[ \frac{2l-1}{l} Q_{l}(x)\right]-l
\big[-Q_{l-1}(x)\big].
\end{equation}
It is now apparent that the terms inside the square brackets can
be absorbed into a redefinition of the characteristic polynomial
through a $x$-independent change of normalization, namely,
\begin{equation}
Q_l(x)\equiv (-1)^l\frac{l!}{(2l-1)!!}P_l(x)=(-1)^l \frac{2^l
(l!)^2}{(2l)!}P_l(x).
\end{equation}
This leads us to the recursion relation of the Legendre
polynomials:
\begin{equation}
(l+1)P_{l+1}(x)= (2l+1)x  P_{l}(x)-l P_{l-1}(x).
\end{equation}

Working along the same lines, it is easy to convince oneself that
the general solution of (\ref{ch-p})  is, up to a normalization
factor, the Jacobi polynomial~$P_l^{a,b}(x)$~\cite{as}:
\begin{equation}\label{jacobi}
Q_l(x)= (-1)^l\frac{2^l l! (l+2m)!}{(2l+2m)!}P_l^{a,b}(x),
\end{equation}
where
\begin{equation}\label{ma-mb}
a=|m_B-m_A|;\qquad b=m_B+m_A ;
\end{equation}
and $m$ is defined in~(\ref{m}). Note that $m$ can be written simply
as~$m=(a+b)/2$.
Note also that $P_{l}^{0,0}$ is the Legendre polynomial $P_{l}$.

>From the result~(\ref{max}) in Appendix~\ref{app}
it turns out that the maximal value of the fidelity~(\ref{fidelity-general-4})
is
attained for $m_A=m_B=0$, i.e., precisely the particular case of Legendre
polynomials discussed above.
Thus, from~(\ref{fidelity-general-4}) we have
\begin{equation}
    F_{\rm max}={1+x_{J+1}^{0,0}\over 2} ,
    \label{optimal fidelity}
\end{equation}
where $x_{n}^{a,b}$ stands for the largest zero of
$P_{n}^{a,b}(x)$. The fact that~$m_A=m_B=0$ implies maximal
fidelity can be translated into physical terms by saying that
Alice's states and Bob's projectors must \emph{effectively } span
the largest possible Hilbert space. For a fixed choice of $m_A$,
not necessarily optimal, the best $m_{B}$ is that for which the
Hilbert spaces spanned by $U(\vec n)\ket{A}$ and $U(\vec
n_{B})\ket{B}$ coincide, i.e., $m_A=m_B=m$. In this case, the
maximal value of the fidelity is given
by~(\ref{fidelity-general-4}), with $x_l=x_{J+1-m}^{0,2 m}$, i.e.,
$F=(1+x_{J+1-m}^{0,2 m})/2<F_{\rm max}$. One reaches the same
conclusion if $m_B$ is fixed and $m_A$ can be adjusted for best
results (see discussion in appendix~\ref{app} after
Eq.~\ref{max}).

\section{General Case: $\Nb$ spins}
\label{sect-general case}

We now show that the solution we have obtained in the preceding
section is in fact of
general validity. Recall that in our original problem
Alice has $N$ spins. Let us suppose that $N$ is even ($N$ odd
will be considered below). As usual, Alice constructs her
states by rotating a fixed eigenstate of~$S_{z}$.
In terms of the irreducible representations of
$SU(2)$, such state can be written as:
\begin{equation}\label{A-repeated}
\ket{A}=\sum_{j=m_A}^{N/2}\left(\sum_\alpha A^{\alpha}_j
\ket{j,m_A;\alpha}\right);\qquad
\sum_{j=m_{A}}^{N/2} \sum_{\alpha} |A_{j}^{\alpha}(m)|^2=1.
\end{equation}
The main difference with the previous example of two equal spins $s$ is that
for
$j<N/2$
the
irreducible representations $U^{(j)}$ appear more than once in the
Clebsch-Gordan decomposition of
$(\mbox{{\boldmath $1/2$}})^{\mbox{\scriptsize{\boldmath $\otimes$}} N}$. Hence,
we label the different occurrences with
the index $\alpha$, which we can view as a new quantum number required
to break the degeneracy of Alice's system of spins under global rotations.
Similarly, the
expression for Bob's  fixed state~$\ket{B}$ is
\begin{equation}\label{B-repeated}
\ket{B}=\sum_{j=m_B}^{N/2}\left(\sum_\beta B^{\beta}_j
\ket{j,m_B,\beta}\right).
\end{equation}
However, it is known that equivalent matrix representations
\begin{equation}
    \Da j \alpha m {m'} (\vec n)=\bra{j,m;\alpha}U(\vec n)\ket{j,m;\alpha}
    \label{equiv}
\end{equation}
are not
orthogonal under the group integration, i.e., for $\alpha\not=\beta$
one has in general
\begin{equation}
 \int dn\;
  \Da  j \alpha m {m'} (\vec n)
 \Dac j \beta m {m'} (\vec n)
 \not=0,
 \label{su2-notpropto}
\end{equation}
and the completeness relation $\int dn_{B}\;O(\vec n_{B})=\mathbb I$ does
{\em not} hold for the simple choice of projectors
$O(\vec n_{B})=c\,U(\vec n_{B})\ket{B}\bra{B}U^\dagger(\vec n_{B})$. We can
circumvent this difficulty by introducing several copies of
$\ket{B}\bra{B}$. A single direction (unit vector) $\vec n_B$ is
thus associated to
\begin{equation}\label{povm-higher}
O(\vec n_{B})=
U(\vec n_B)\;\Big[\;\ket{B}\bra{B}+\ket{B'}\bra{B'}+\ket{B''}\bra{B''}+
\ \cdots \Big]\;U^{\dagger}(\vec n_B).
\end{equation}
The fixed projectors in the square brackets will be judiciously
chosen to eliminate the off-diagonal terms
coming from the mixing
of equivalent representations in the closure relation.
The projector $O(\vec n_{B})$ are explicitly of rank higher than
one. However, recalling~\cite{davies}, we can view the right hand side
of~(\ref{povm-higher}) as defining a sum of rank one projectors
$O(\vec n_{B})+O'(\vec n_{B})+O''(\vec n_{B})+\cdots$. The two points of
view are equivalent if the averaged fidelity is used as a
figure of merit.
In a suggestive compact notation
we can write
\begin{equation}
\ket{B}\bra{B}+\ket{B'}\bra{B'}+\ket{B''}\bra{B''}+~\cdots\  \equiv
\ket{\Bb}\cdot\bra{\Bb},
\end{equation}
where
\begin{equation}
\ket{\Bb}\equiv \sum_{j=m_B}^{N/2}\left(\sum_\beta \Bb^{\beta}_j
\ket{j,m_B,\beta}\right),
\end{equation}
and
\begin{equation}
\Bb_j^\beta\equiv (B_j^\beta,{B'}_j^{\beta},{B''}_{j}^\beta,\dots) .
\end{equation}
Next, we introduce a set of orthonormal vectors $\{\bb_{j}^\alpha\}$;
\begin{equation}
\label{bxb}
\bb_j^\alpha \cdot \bb_j^\beta =\delta^{\alpha\beta},
\end{equation}
and define the vectors $\Bb_j^\alpha$ as
\begin{equation}
\Bb_j^\alpha= \sqrt{\frac{2j+1}{c}}\;\bb_j^\alpha.
\end{equation}

With the above definitions one can easily
see that $\int dn_{B}\;O(\vec n_{B})=\mathbb I$ and, hence, the set of
projectors~(\ref{povm-higher}) defines a POVM.

The fidelity can be read off from~(\ref{fidelity-general-2}) and it is
given by
\begin{equation}\label{fidelity-rep}
F=\frac{1}{2}+\frac{1}{2}\sum_{j=m}^{N/2} \sum_{\alpha} \mu_j (A_j^\alpha)^2 +
\sum_{j=m+1}^{N/2}\sum_{\alpha\beta}
A_{j-1}^\alpha\left(\bb_{j-1}^\alpha\cdot\bb_{j}^\beta\right) A_{j}^\beta
\nu_j-{1\over2}\sum_{j=m_A}^{m-1}\sum_{\alpha} (A_j^\alpha)^2,
\end{equation}
where the phases have been chosen so that $\nu_{j}$, $A_{j}^\alpha$
and $B_{j}^\alpha$
are real.
In general $\bb_{j}^\alpha\in\mathbb R^k$, where $k$ must be greater or
equal than the highest degeneracy of the irreducible representations
in the Clebsch-Gordan series of
$(\mbox{{\boldmath $1/2$}})^{\mbox{\scriptsize{\boldmath $\otimes$}}
N}$, since  otherwise~(\ref{bxb}) could not be fulfilled.
Equation~\ref{fidelity-rep} suggests the definition
\begin{equation}
    \Ab_{j}=\sum_{\alpha} A_{j}^\alpha \bb_{j}^\alpha,
    \label{Avec}
\end{equation}
which enables us to write
\begin{equation}\label{fidelity-rep2}
F=\frac{1}{2}+\frac{1}{2}\sum_{j=m}^{N/2}  \mu_j |\Ab_j|^2 +
\sum_{j=m+1}^{N/2}
\Ab_{j-1}\cdot\Ab_{j}
\nu_j-{1\over2}\sum_{j=m_A}^{m-1} |\Ab_j|^2.
\end{equation}
Using Schwarz inequality we have
\begin{equation}\label{fidelity-rep3}
F\le \frac{1}{2}+\frac{1}{2}\sum_{j=m}^{N/2}  \mu_j |\Ab_j|^2 +
\sum_{j=m+1}^{N/2}
|\Ab_{j-1}||\Ab_{j}|
\nu_j-{1\over2}\sum_{j=m_A}^{m-1} |\Ab_j|^2.
\end{equation}
The right hand side is exactly the fidelity~(\ref{fidelity-general-2})
of the preceding section with the substitution
\begin{equation}
    A_{j}\to\tilde{A}_{j}\equiv
    |\Ab_{j}|= \sqrt{\sum_{\alpha}(A_{j}^\alpha)^2}.
    \label{mod}
\end{equation}
This equation shows that the existence of several
equivalent representations in the Clebsch-Gordan decomposition of
Alice's Hilbert space
cannot be used to increase the value of the fidelity already obtained in
section~\ref{sect-2 spins s}. The equality holds when all
vectors $\Ab_{j}$ are parallel, in which case we
recover~(\ref{fidelity-general-2}). The square root on the right hand
side of~(\ref{mod}) plays the role of an effective component
of $\ket{A}$ on the Hilbert space of a {\em single}
irreducible representation {\boldmath $j$}. The specific ways $\ket{A}$
projects on
each one of the equivalent representations
are of no relevance, provided $\tilde{A}_{j}$ do not change. As far as the fidelity is concerned, all them are
equivalent to taking a state
$\ket{\tilde A}$
that belongs to {\boldmath $N/2\oplus (N/2-1)\oplus
(N/2-2)\oplus\cdots$} (no duplications),
with the corresponding components given by~$\tilde{A}_{j}$.

\bigskip

As we have just seen, the maximal fidelity can be achieved from a
code state containing only one of each irreducible
representations. This type of states are formally the same as those
considered in the simplified example  of two equal spins
$s_1=s_2=s$ studied in section~\ref{sect-2 spins s}, for which
{\boldmath $s\otimes s= J\oplus
(J-1)\oplus\cdots\oplus
0$}, with $J=2s=N/2$. The problem of an even number of spins is
thus completely solved: according to~(\ref{fidelity-general-4})
the maximal fidelity is given by
\begin{equation}
F_{N}={1+x^{0,0}_{N/2+1}\over2},\qquad  \mbox{for $N$ even},
    \label{xl even}
\end{equation}
where~$x^{0,0}_{N/2+1}$ is the largest
zero of the (Legendre) polynomial $P_{N/2+1}(x)=P^{0,0}_{N/2+1}(x)$.

For an odd number of spins we can proceed as in
section~\ref{sect-2 spins s} but considering now states with two
different spins: $s_{1}=s$, $s_{2}=s+1/2$. The corresponding
Clebsch-Gordan decomposition is also non-degenerate: {\boldmath
$s \otimes(s+1/2)= J\oplus (J-1)\oplus\cdots \oplus 1/2$}, with
$J=2s+1/2=N/2$. The results from~(\ref{A-general})
to~(\ref{ch-p}) are still valid (for the value of $J$ we have
just specified). The optimal values of~$m_{A}$ and~$m_{B}$ are
again the minimal ones: $m_A=m_B=1/2$. The maximal fidelity is
\begin{equation}
F_{N}={1+x^{0,1}_{N/2+1/2}\over2},\qquad  \mbox{for $N$ odd},
    \label{xl odd}
\end{equation}
where $x^{0,1}_{N/2+1/2}$ stands for the
largest zero of the Jacobi polynomial $P_{N/2+1/2}^{0,1}(x)$. This
completes the solution of the general problem.

It is physically obvious that the larger the number of spins
Alice can use the better she should be able to encode $\vec n$.
One thus expects that the maximal fidelity should increase
monotonously with $N$. It is interesting to obtain this result
from the properties of the zeroes of the Jacobi polynomials.  For
an even number of spins, $N=2n-2$, the corresponding zero is
$x_{n}^{0,0}$, whereas for $N+1$ it is $x_{n}^{0,1}$, and
$x_{n-1}^{0,1}$ for $N-1$. Proving that $F_{N-1}<F_{N}<F_{N+1}$
amounts to showing that
\begin{equation}\label{mon}
x_{n-1}^{0,1}<x_{n}^{0,0}<x_{n}^{0,1},
\end{equation}
but this is just a particular case of~(\ref{b+1}) for $a=0$ and $b=1$.

Not only the optimal strategy Alice can devise with $N$ spins leads to a
fidelity
larger than $F_{N-1}$.
She can also use non-optimal
ones and still exceed $F_{N-1}$.
E.g., for $N=4$, the choice
$m_A=m_B=1$, which is non-optimal, gives a fidelity
$F= (10+\sqrt{10})/15 >(6+\sqrt{6})/10=F_{3}$. This is
also a trivial consequence
of~(\ref{b+1}) as in this case one has $x_2^{0,2}>x_2^{0,1}$. In
physical terms, this is telling us that the dimension of the
Hilbert space spanned by $U(\vec n)\ket{A}$ and $U(\vec n_{B})\ket{B}$
when $N=4$ and  $m_A=m_B=1$ (including equivalent spin representations
only once) is still larger
than the maximal available dimension for $N=3$.

\section{Discussion and outlook}\label{sect-results and discussion}

In this paper we have addressed the problem of optimizing
strategies for encoding and decoding directions on
the quantum states of a system of $N$ spins. We have restricted
ourselves to states that point along a definite direction in an
intrinsic way, namely, to eigenstates of $\vec n\cdot \vec S$. This
case is of great interest since no prior knowledge of any
sender's (Alice's) reference state or frame by the recipient (Bob)
is required at all for
a viable transfer of the information.
We have optimized both Alice's states and Bob's measurements. Our
results are summarized in~(\ref{xl even}) and~(\ref{xl odd}),
where we give the maximal averaged fidelities $F_{N}$.
Interestingly enough, these results can be written in terms
of the largest zeroes of the Jacobi polynomial, which are known to
play an important role in angular momentum theory and are intimately related
to the matrix representations of $SU(2)$. The states that lead to the
maximal fidelities are among those that have the smallest (non-negative) value
of~$\vec
n\cdot\vec S$, namely, $m=0$ for $N$ even and $m=1/2$ for $N$ odd, but
still span the largest Hilbert space under rotations.

We display the values of the maximal fidelity
for $N$ up to seven in table~\ref{table-I} for illustrational purposes.
\begin{table}
\begin{center}
\begin{tabular}{c|ccccccc}
  \toprule
  $N$ & 1 & 2 & 3 & 4 & 5 & 6 & 7 \\
  \colrule
  $F_{N}$ & $\frac{2}{3}$ & $\frac{3+\sqrt{3}}{6}$ & $\frac{6+\sqrt{6}}{10}$
&
  $\frac{5+\sqrt{15}}{10}$
  & $.9114$ & $.9306$ & $.9429\phantom{\Big|}$\\
  \botrule
\end{tabular}
\end{center}
\caption{Maximal fidelities as a function of the number of
spins.\label{table-I}}
\end{table}
It shows, e.g., that the optimal encoding with three spins
($m=1/2$) gives
$F_{3}=(6+\sqrt{6})/10 \sim 0.845 $, which is  already larger than
the corresponding maximal value for four {\em parallel} spins ($m=2$):
$F=5/6\sim .833$
\cite{mp}. This illustrates a general feature: the optimal strategies
discussed here lead to fidelities that increase with~$N$ much faster than
that of sending parallel spins.
In fact, Eq.~\ref{asy} shows that $F_{N}$ approaches unity
quadratically in the number of spins, namely
\begin{equation}\label{assymptotic-1}
  F_{N}\sim 1-\frac{\xi^2}{ N^2},
\end{equation}
where $\xi\sim 2.4$ is the first zero of the Bessel
function~$J_0(x)$. In contrast, if parallel spins are used
the maximal fidelity approaches
unity only linearly, $F\sim 1-1/{N}$.

This can be understood in terms of the dimension $d$
of the Hilbert space used {\em effectively} in each case, which
is a direct sum of the Hilbert spaces of the irreducible
representations of $SU(2)$ involved. Here {\em
effectively} means `non-redundantly', thus equivalent representations
count only once.
Encoding with $N$ parallel spins uses only the Hilbert space of the
representation~{\boldmath $N/2$}, whose dimension is
$d=N+1$, whereas our optimal strategy uses a much larger Hilbert space,
with $d=(N/2+1)^2$ for $N$
even and $d=(N/2+1)^2-1/4$ for $N$ odd; in both cases $d\sim N^2$.
We are led to the conclusion that the fidelity as a function of $d$
tends to unity as
\begin{equation}
F\sim 1-{a\over d} ,
\label{F(d)}
\end{equation}
where $a$ is of order one and
depends on the particular strategy.

Improvements on the approach discussed in this paper can only come
from encoding and decoding procedures that make extensive use of
the available Hilbert space, namely,  strategies that use the
redundant equivalent representations. In~\cite{us} we presented a
strategy for which the maximal fidelity  approaches unity
exponentially in the number of spins, i.e.,  $F\sim1-2^{-N}$. We
argue there that this encoding is likely to lead to the maximal
fidelity one can possibly achieve with $N$ spins, since it makes
effective use of the whole Hilbert space of the system, for which
$d=2^{N}$ (thus, Eq.~\ref{F(d)} also holds in this case). The
corresponding encoding process, however, involves complicated
unitary operations and, moreover, it seems to require that Alice
and Bob share a common reference frame~\cite{peres-private}.

We have obtained our general results
using
continuous POVMs, but finite ones can also be designed.
For $N$ parallel spins ($m_A=m_B=N/2$), a general recipe for
finite optimal POVMs exists~\cite{derka}, and minimal versions
for up to $N=7$ can be
found in~\cite{lpt}. The unit vectors $\vec n_{r}$ associated to the
outcomes of these POVMs
are the vertices of certain
polyhedra inscribed in the unit sphere. For $N\leq 7$ we have explicitly
verified that these very same polyhedra can be use to
design finite
optimal POVMs for any value of $m_{A}=m_{B}\leq N/2$.
Moreover, the minimal POVMs of~\cite{lpt} remain minimal for the
states considered here. We have discussed this issue in detail
for~$N=2$
in section~\ref{sect-2 spins}.
For $N=3$ the polyhedron corresponding to the minimal POVM is the
octahedron~\cite{lpt}. One can easily verify that $O_{r}=U(\vec n_{r})
\ket{B}\bra{B}U^\dagger(\vec n_{r})$, where $\ket{B}$ is given
in~(\ref{B-general}) with $m_{B}=1/2,\, 3/2$, fulfill the completeness
condition~(\ref{povm}).
We hence believe that the discretization of a
continuous POVM is a geometrical problem, i.e., it seems to be
independent of the states $\ket{B}$.

The optimal states, $\ket{A}$, can be easily computed from the
matrix $\mathsf M$ in~(\ref{matrix}),  as they are the eigenvectors
corresponding to the
maximal eigenvalue. Recall that for $N=2$ one obtains the
one-parameter family of states~(\ref{eigenvector-2}) which includes
the product states
$\ket{\uparrow\downarrow}$, $\ket{\downarrow\uparrow}$.
For $N>2$, product states of the type
$\ket{\uparrow\downarrow\uparrow\uparrow\downarrow \cdots}$ do not
seem to be optimal. Consider, e.g.,
$N=4$. The optimal eigenvector of $\mathsf M$ is
\begin{equation}
\ket{A}=\frac{\sqrt{2}}{3}\ket{2,0}+e^{i\gamma_1}\frac{1}{\sqrt{2}}\ket{1,0}
+e^{i\gamma_0}\sqrt{\frac{5}{18}}\ket{0,0},
\label{eigenvector}
\end{equation}
which is clearly not a product state of the individual spins for
any choice of the phases~\cite{1+1}.
One could argue that this solution is not entirely
general because the Clebsch-Gordan series
of~$(\mbox{{\boldmath $1/2$}})^{\mbox{\scriptsize{\boldmath $\otimes$}}
4}$ contains the
representation $\mathbf 1$ three
times and $\mathbf 0$ twice, whereas in~(\ref{eigenvector}) they
appear only once.
However, {\em any} optimal state has the same
`effective' components $\tilde A_{j}$
(see eqs.~\ref{fidelity-rep3}-\ref{mod}),
which can be read off from~(\ref{eigenvector}):
\begin{equation}
\tilde A_{2}={\sqrt{2}\over3},\qquad \tilde A_{1}={1\over\sqrt{2}},\qquad
\tilde A_{0}=\sqrt{{5\over18}}.
\label{eigenvector tilde}
\end{equation}
Note now that {\em any} product state with~$m=0$ (two spins up and two spins
down), e.g.,
$\ket{\uparrow\uparrow\downarrow\downarrow}$,
$\ket{\uparrow\downarrow\downarrow\uparrow}$, has an
`effective'
Clebsch-Gordan decomposition given by $\tilde A_{2}=\tilde A_{1}=\tilde
A_{0}=1/\sqrt{3}$, which do not coincide with~(\ref{eigenvector
tilde}). Therefore, these product states cannot be optimal.
Nevertheless, they lead to a maximal fidelity
$F=(15+5\sqrt{2}+2\sqrt{5})/30\approx
0.885$, which is remarkably close to
$F_{4}\approx0.887$.
This is likely to be the case for arbitrary~$N$. These issues
are currently under investigation.

\section*{Acknowledgments}
We thank S.~Popescu, A.~Bramon, G.~Vidal and W.~D\"ur for
stimulating discussions. Financial support from CICYT contracts
AEN98-0431, AEN99-0766, CIRIT contracts 1998SGR-00026,
1998SGR-00051, 1999SGR-00097 and EC contract IST-1999-11053 is
acknowledged.

\appendix
\section{ }\label{app}

In this appendix we collect the mathematical properties of the Jacobi
polynomials $P_{n}^{a,b}(x)$ that we use in the text. We are concerned
only with integer values of $a$ and~$b$ such that $b\geq a\geq 0$.
Further properties can
be found in~\cite{as} and~\cite{nu}.

For fixed $a$ and $b$, $\{P_{n}^{a,b}(x)\}$ is a set of
orthogonal
polynomials, where~$n$ labels the degree of each polynomial in the set.
A convenient definition can be stated in terms of
their Rodrigues formula:
\begin{equation}\label{j-def}
P_{n}^{a,b}(x)=\frac{(-1)^n}{2^n n!} (1-x)^{-a}(1+x)^{-b}
\frac{d^n}{{dx}^n} \left[ (1-x)^{n+a}(1+x)^{n+b}\right].
\end{equation}
From~(\ref{j-def}) it follows the recursion relation:
\begin{equation}\label{j-rec}
xP_{n}^{a,b}(x)=\alpha_n P_{n+1}^{a,b}(x)+\beta_n
P_{n}^{a,b}(x)+\gamma_n P_{n-1}^{a,b}(x),
\end{equation}
with
\begin{eqnarray}
\alpha_n&=& \frac{2(n+1)(n+a+b+1)}{(2n+a+b+1)(2n+a+b+2)},\nonumber \\
\beta_n &=&\frac{b^2-a^2}{(2n+a+b)(2n+a+b+2)},\nonumber\\
\gamma_n &=&\frac{2(n+a)(n+b)}{(2n+a+b)(2n+a+b+1)}.
\end{eqnarray}
It also follows the differentiation formula
\begin{equation}\label{j-dif}
 \frac{d\;P_{n}^{a,b}(x)}{dx}=\frac{n+a+b+1}{2}P_{n-1}^{a+1,b+1}(x).
\end{equation}
The normalization is chosen so that the coefficient $A_n$ of the
highest power of $P_{n}^{a,b}(x)=A_n x^n+B_n x^{n-1}+\cdots$ is
\begin{equation}\label{j-norm}
  A_n=\frac{\Gamma(2n+a+b+1)}{ 2^n n! \Gamma(n+a+b+1)} .
\end{equation}

The following two relations can also be obtained from the
definition~\ref{j-def}:
\begin{eqnarray}\label{b-rel-1}
 (2n+a+b) P_n^{a,b-1}(x)&=&(n+a+b) P_n^{a,b}(x)+(n+a)
 P_{n-1}^{a,b}(x),\\
\label{b-rel-2}
 (n+b+a+1)\frac{1+x}{2} P_n^{a,b+1}(x)&=&
 (n+1) P_{n+1}^{a,b-1}(x)+b P_{n}^{a,b}(x).
\end{eqnarray}

Let us recall some basic facts about the zeros of orthogonal
polynomials: {\it i}) any orthogonal polynomial, $P_n$, of order
$n$ has~$n$ real simple zeros. For Jacobi polynomials these zeros
lie in the interval $(-1,1)$; {\it ii}) the zeros of $P_n$ and
$P_{n+1}$ are interlaced; {\it iii})  for $x$ greater than the
largest zero, the polynomial is a monotonously increasing function
(if the polynomial is normalized as in Eq.~\ref{j-norm}, where
$A_{n}>0$). In
particular, $P_n(x)$ must be positive in this region .

Now we can prove the results needed in the text. As in there, we denote by
$x_n^{a,b}$ the largest zero of the polynomial $P_{n}^{a,b}(x)$.
Let us start by showing that
 \begin{equation}\label{a+1}
  x_{n-1}^{a+1,b+1}<x_n ^{a,b}.
\end{equation}
From property~{\it iii} above it follows that the left hand side
of~(\ref{j-dif}) is manifestly positive for $x>x_n^{a,b}$. Hence,
so it is the right hand side. We conclude that
$x_{n-1}^{a+1,b+1}$ cannot belong to this region and~(\ref{a+1})
follows~$\blacksquare$

Next, we prove the inequality
\begin{equation}\label{b+1}
x_{n-1}^{a,b}<x_{n}^{a,b-1}<x_{n}^{a,b}.
\end{equation}
We evaluate~(\ref{b-rel-1}) at
$x=x_n^{a,b}$ and use properties~{\it ii} ($\Rightarrow x_{n-1}^{a,b}<x_n^{a,b}$)
and~{\it iii}, which imply
that $P_{n-1}^{a,b}(x_n^{a,b})>0$, to obtain that $P_n^{a,b-1}(x_n^{a,b})>0$.
We repeat the process for $x=x_{n-1}^{a,b}$ and conclude
that $P_n^{a,b-1}(x_{n-1}^{a,b})<0$. Hence $P_n^{a,b-1}$ has a zero in the
interval $(x_{n-1}^{a,b},x_{n}^{a,b})$. This is necessarily the
largest zero $x_n^{a,b-1}$ since, according to~(\ref{b-rel-1}) and
properties~{\em ii} and~{\em iii},
$P_n^{a,b-1}(x)>0$ for $x>x_n^{a,b}$. Thus~(\ref{b+1})
follows~$\blacksquare$

The inequality
\begin{equation}\label{b+2}
x_{n}^{a,b+1}<x_{n+1}^{a,b-1}
\end{equation}
can be proven as follows.
Evaluate~(\ref{b-rel-2}) at $x=x_n^{a,b+1}$ so that the left hand
side of this equation is zero. The second
inequality in~(\ref{b+1}) and property {\it iii} imply that
$P_{n}^{a,b}(x_{n}^{a,b+1})>0$. Hence the first term on the right hand
side of~(\ref{b-rel-2}) must be negative, i.e.,
$P_{n+1}^{a,b-1}(x_{n}^{a,b+1})<0$, and~(\ref{b+2}) follows
immediately, since otherwise property~{\em iii} would not hold for
$P_{n+1}^{a,b-1}$~$\blacksquare$

For two given integers $l,m$ consider now the following  set of zeros
\begin{equation}
C_{m}^l=
\{ x_{l-m''}^{m''-m',m''+m'}:\, m\leq m'
\leq m''\leq l \}.
\label{clm}
\end{equation}
We want to prove that
\begin{equation}
\max{C_{m}^l}=x_{l-m}^{0,2m}. \label{max}
\end{equation}
According to~(\ref{a+1}), decreasing~$m''$ by one leads us
to a larger zero. The maximum is then in the subset $\{
x_{l-m'}^{0,2 m'} :  m\leq m'\leq l \}$. The inequality~(\ref{b+2})
now implies~(\ref{max})~$\blacksquare$

Finally, we give the large $n$ (asymptotic) behavior
of~$x_n^{a,b}$~\cite{as}:
\begin{equation}\label{asy}
  x_n^{a,b}= 1-\frac{\xi_{a}^2}{2 n^2}+O(\frac{1}{n^3}),
\end{equation}
where $\xi_{a}$ is the first zero of the Bessel function
$J_a(x)$.
For $a=0$, which is relevant for our discussion in
section~\ref{sect-results and discussion}, we
also give the subleading term:
\begin{equation}
x_n^{0,b}= 1-\frac{\xi_{0}^2}{2 n^2}\left( 1-\frac{b+1}{n}
\right)+O(\frac{1}{n^4}),
\end{equation}
where
\begin{equation}\label{xi}
   \xi_{0}=\xi=2.405.
\end{equation}


\begin{thebibliography}{99}
\bibitem{pw} A. Peres and W. K. Wootters,
             Phys. Rev. Lett. {\bf 66}, 1119 (1991).
\bibitem{mp} S. Massar, Phys. Rev.
             Lett. {\bf 74}, 1259 (1995) .
\bibitem{derka} R. Derka, V. Buzek, A.K. Ekert,
             Phys. Rev. Lett.\ {\bf 80}, 1571 (1998).
\bibitem{lpt}J.I. Latorre, P. Pascual, R. Tarrach,
             Phys. Rev. Lett. {\bf 81}, 1351 (1998).
\bibitem{gp} N. Gisin and S. Popescu,
             Phys. Rev. Lett. {\bf 83}, 432 (1999).
\bibitem{us} E.Bagan {\em et al.}, Phys. Rev. Lett.~{\bf 85}, 5230
             (2000); quant-ph/0006075.
\bibitem{ps} A. Peres and P. Scudo, quant-ph/00010085.
\bibitem{massar} S. Massar, Phys. Rev. {\bf A62} (2000) 040101(R).
\bibitem{davies}E.B. Davies,
             IEEE Trans. Inform. Theory IT-24, 596 (1978).
\bibitem{holevo} A. S. Holevo, {\it Probabilistic and Statistical
             Aspects of Quantum Theory}, North Holland, Amsterdam, 1982.
\bibitem{wktung}W.K. Tung, {\it Group Theory in Physics}
             World Scientific Publishing, 1985;
             A.R. Edmonds, {\it Angular Momentum in Quantum Mechanics},
             Princeton University Press, 1960.
\bibitem{as} For standard definitions and conventions see, for instance:
             M. Abramowitz and I.A. Stegun, {\it Handbook of
             Mathematical
             Functions}, Dover, New York 1970.
\bibitem{peres-private} A. Peres, private communication.
\bibitem{1+1} If we consider this state as a bipartite system of two spin~1
             subsystems, it is also entangled for any choice of the
             phases $\gamma_{1}$, $\gamma_{0}$.
\bibitem{nu} A.F. Nikiforov and V.B. Uvarov,
             {\it Special Functions of Mathematical
             Physics}, Birkh\"{a}suer, Basel, 1988.

\end{thebibliography}
\end{document}